\documentclass[11pt]{article}
\usepackage{graphicx}
\usepackage{amsmath}
\begin{document}
\begin{center}
{\Large\bf  Azimuthal Dependence of DIS with Spin-1 Target }
\par\vskip20pt
J.P. Ma$^{1,2}$, C. Wang$^{3}$ and  G.P. Zhang$^{3}$     \\
{\small {\it
$^1$ Institute of Theoretical Physics, Academia Sinica,
P.O. Box 2735,
Beijing 100190, China\\
$^2$ Center for High-Energy Physics, Peking University, Beijing 100871, China  \\
$^3$ School of Physics,  Peking University, Beijing 100871, China
}} \\
\end{center}
\vskip 1cm
\begin{abstract}
We study DIS with a spin-1 hadron or a nucleus polarized 
in arbitrary direction. The differential cross-section in this case will have an azimuthal dependence. 
We derive the dependence by including contributions from twist-2- and twist-3 QCD operators. The twist-3 contribution is computed at tree-level.
A spin-1 hadron or nucleus can have nonzero tensor polarization. 
 We find that all structure functions related to the tensor-polarization 
of the initial hadron can be extracted by study the azimuthal dependence. A nonzeo result of the structure functions 
of a nucleus from experiment will indicate nontrivial inner structure of the nucleus. 

\vskip 5mm
\noindent
% PACS numbers:
\end{abstract}

% Re-write the paper starting on 3. Dec. 2013 to address the problem with frames. 

\vskip 1cm

\par\vskip20pt
\par
DIS experiment has been done mostly with a proton target. It has provided a wealth of information about the 
inner structure of proton and has played important role in testing QCD as the theory of strong interaction. 
If one takes a nucleus target as a weakly bounded state, one may expect 
that DIS cross-section is a sum of DIS cross-sections of each nucleon in the nucleus. 
But, EMC experiment has shown that it is not the case\cite{EMC}. In fact, the interaction 
between nucleons in a nucleus can not be neglected, the nucleons are correlated. This is clearly shown 
by the measurement in \cite{SRN} of DIS cross-sections with Bjorken variable of a nucleon larger than one. It implies 
that in DIS scattering with a nucleus more than one nucleon is involved.   
\par
DIS with a spin-1 target like a spin-1 nucleus has been theoretically analyzed in \cite{HJM}.     
There are more structure functions than those in DIS with a proton. Especially, there are  structure functions
related the tensor polarization of the target. These functions do not exist for a proton target. It has been shown in \cite{HJM} 
that these structure functions are zero, if nucleons inside the nucleus are at rest and 
do not interact with each other. In a recent experiment, one has found the evidence that one of these structure functions
of deuteron, 
called as $b_1$, is nonzero\cite{b1hermes}. It is clear that progresses made in experimental study of DIS with a nucleus will help 
to explore interactions or correlations between 
nucleons in a nucleus, and partonic picture of a nucleus.   

\par 
In this work we study DIS with a spin-1 target. We consider the case that the initial hadron is polarized in an arbitrary 
direction. In this case the differential cross-section will have an azimuthal dependence. We will derive this dependence. 
Measuring the dependence 
will help to disentangle various structure functions. Our analysis is made in the framework of QCD factorization. We include 
not only contributions of twist-2 operators but also twist-3 operators. 
We re-analyze contributions of twist-2 operators and confirm the existing results with twist-2 operators. 
At twist-3 there are two additional distributions contributing to structure functions at tree-level. Again, these distributions 
can be extracted from the azimuthal dependence.  
\par 
We consider the process: 
\begin{equation} 
 \ell + H  \to \ell + \gamma^* + H \to 
 \ell  +X  , 
\end{equation}
where $H$ is a spin-1 hadron. We will consider the case in which $H$ is polarized along an arbitrary direction and the polarization is described by a polarization vector $\epsilon$ with $\epsilon\cdot P=0$, where $P$ is the momentum of $H$.  We take a frame in which $H$ moves in the $z$-direction 
with the momentum $P$.  The initial hadron $H$ is polarized. The initial lepton carries the momentum $k$ and is polarized 
with the helicity $\lambda$.

\par
The hadronic tensor is defined as:
\begin{equation}
    W^{\mu\nu}=\frac{1}{4\pi} \sum_{X}\int d^4 x e^{iq\cdot x}
    \langle H \vert J^{\mu}(x) \vert X  \rangle
    \langle X  |J^{\nu}(0) \vert H  \rangle. 
\label{WT}     
\end{equation}
The relevant  standard variables are: 
\begin{equation}  
x_B =-\frac{q^2}{2 q\cdot P}=\frac{Q^2}{2q\cdot P}, \ \ \   y=\frac{q\cdot P}{P\cdot k}.  
\end{equation}
The Bjorken variable $x_B$ is in the range $ 0< x_B <1$. In the case that the initial hadron is a nucleus with the atomic number 
$A$, one can define the Bjorken variable $x_b$ of a nucleon with the relation $x_b = A x_B$. As mentioned 
at the beginning, the DIS cross-section with a nucleus is not zero for $x_b >1$.    
We assume that the initial hadron is polarized along the direction $\vec m$ in the sense that the projection of the spin vector 
along $\vec m$ is quantized, i.e., $ \vec m \cdot \vec S = \lambda_m =0,\pm 1$. 
We introduce a frame in which the initial lepton moves in 
the $-z$-direction and the initial hadron in the $+z$-direction. The $x$-axis of the frame is chosen 
so that $\vec m$ is in the $xz$-plan with the angle $\theta_m$ to the $z$-direction. In this frame the lepton in the final state moves 
in the direction described with the polar angle $\theta$ and the azimuthal angle $\phi$. We call this frame as laboratory frame.  
This is illustrated in Fig.1. 
\par 
\begin{figure}[hbt]
\begin{center}
\includegraphics[width=8cm]{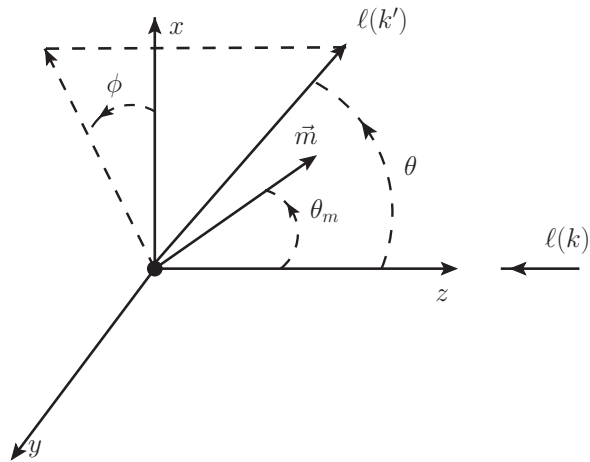}
\end{center}
\caption{The defined frame for the differential cross-section.    }
\label{P1}
\end{figure}

\par 
In the framework of collinear factorization the hadronic tensor can be factorized as convolutions
of perturbative coefficient functions with various matrix elements of QCD operators at different 
twists. In this work we only consider the contributions from twist-2- and twist-3 QCD operators. 
Therefore we write the hadronic tensor as:
\begin{equation} 
   W^{\mu\nu} = W^{\mu\nu}\biggr\vert_{{\rm twist}-2} + W^{\mu\nu}\biggr\vert_{{\rm twist}-3} +
   \cdots. 
\label{TWW}   
\end{equation} 
The first term only receives the contributions from twist-2 operators in collinear factorization and the leading term in the Bjorken limit, the second term receives contributions only from twist-3 operators and is relatively suppressed by 
$Q^{-1}$. The remaining terms denoted by $\cdots$ are of higher twists are suppressed by $Q^{-n}$ with $n \ge  2$ relative to the first term.     
In the separation of Eq.(\ref{TWW}) we neglect the effect of the hadron mass $M$ except the effect in the relation between 
$x_B$ and the momentum fraction $x$ of partons. The neglected effect are suppressed at least by $M^2/Q^2$ in our results. 

\par
The hadronic tensor is conveniently analyzed in the light-cone frame in which the virtual photon moves in the $-z$-direction and the hadron in the $z$-direction. 
In this frame a vector
$a^\mu$ is expressed as $a^\mu = (a^+, a^-, \vec a_\perp) =
((a^0+a^3)/\sqrt{2}, (a^0-a^3)/\sqrt{2}, a^1, a^2)$ and $\vec a_\perp^2
=(a^1)^2+(a^2)^2$. We introduce two light-cone vectors 
$l^\mu=(1,0,0,0)$ and $n^\mu=(0,1,0,0)$. Two tensors related to the two vectors are defined as: $g_\perp^{\mu\nu} = g^{\mu\nu} - n^\mu l^\nu - n^\nu l^\mu $ and $ \epsilon_\perp^{\mu\nu} =\epsilon^{\alpha\beta\mu\nu}l_\alpha n_\beta$. In the frame the momentum $P$ is $P^\mu =(P^+,P^-,0,0)$ with $P^2=M^2$ and the virtual photon carries the momentum $q^\mu=(q^+,q^-,0,0)$. The introduced frame is not equivalent to the laboratory frame in Fig.1. 
However, the obtained hadronic tensor will be given in a covariant form by noting that $g_\perp^{\mu\nu}$ and $\epsilon_\perp^{\mu\nu}$ can be defined 
in terms of $g^{\mu\nu}$, $P^\mu$ and $q^\mu$ in an arbitrary frame. From any vector $a^\mu$ one can project the vector $a_\perp^\mu = g_\perp^{\mu\nu}a_\nu$, which is transverse to $P$ and $q$. 

\par 
With the polarization vector $\epsilon^\mu$ one can define 
the spin density matrix as $\rho^{\mu\nu} \propto \epsilon^\mu \epsilon^{*\nu}$. Instead of using polarization vectors, it will be convenient 
to use the following quantities to describe the spin. Although we explicitly define them in the light-cone frame, 
these quantities can also be defined covariantely as discussed in the above. 
From the symmetric- and trace-less part we introduce:
\begin{eqnarray} 
  && T_\perp^{\mu\nu} = \frac{1}{2} \biggr ( \epsilon_\perp^\mu \epsilon_\perp^{*\nu} +  \epsilon_\perp^\nu \epsilon_\perp^{*\mu} 
   - g_\perp^{\mu\nu} \epsilon_\perp \cdot \epsilon_\perp^* \biggr ),      
\nonumber\\
  && \tilde S_{\perp}^\mu = \frac{M}{P^+} \biggr ( n\cdot \epsilon^* \epsilon_\perp^\mu +  n\cdot \epsilon \epsilon_\perp^{*\mu} \biggr ), 
  \ \ \ \  \tilde S_L  = -\frac{1}{2} - \frac{3}{2} \biggr ( n\cdot \epsilon l\cdot \epsilon^* + l\cdot \epsilon n \cdot \epsilon^* \biggr ), 
\end{eqnarray} 
and from the anti-symmetric part of the spin-density matrix
one can define a transverse vector and the helicity constant $S_L$:
\begin{eqnarray}  
       S_\perp^\mu =-i \epsilon_\perp^{\mu\nu}\frac{M}{P^+} \biggr ( n \cdot \epsilon \epsilon^{*}_{\perp\nu}
          -n\cdot \epsilon^* \epsilon_{\perp\nu} \biggr ), \ \ \ 
     S_L =  i \biggr ( \epsilon^1 \epsilon^{*2}- \epsilon^2 \epsilon^{*1} \biggr ).          
\end{eqnarray} 
The quantities introduced from the symmetric- and trace-less  part of $\rho^{\mu\nu}$ describe 
the tensor polarization of $H$. If there is no tensor polarization, all of these quantities are zero. 
Those from the antisymmetric part describe the vector 
polarization. It is easy to find that $S_L$ is the helicity.  Comparing the spin description 
of a spin-1/2 hadron, where one only needs a helicity and a transverse-spin vector for helicity-flip, 
corresponding to  $S_L$ and $S_\perp^\mu$, respectively. Interpretations of the quantities introduced in the above 
are discussed in detail in \cite{BaMu}, where production of a spin-1 hadron in semi-inclusive DIS is studied.     
\par

The twist-2 contributions have been derived in \cite{HJM,JaMa} with the technique of operator product expansion(OPE). We re-derive these results and organize the results 
in the way that the one-loop corrections of $\alpha_s$ can be easily obtained from existing results of DIS with spin-1/2 target. 
The relevant matrix element of twist-2 quark operator can be 
parametrized as:
\begin{equation} 
\int\frac{d\lambda}{2\pi} e^{-ix\lambda P^+} 
  \langle H  \vert \bar \psi_i  (\lambda n)   \psi_j (0) \vert H \rangle =  \frac{1}{2} 
  \biggr [  \left ( q(x) +  \tilde S_L \tilde q_{L} (x)  \right ) \gamma^- + S_L q_L (x) \gamma_5 \gamma^- \biggr ]_{ji} + \cdots , 
\label{Quark} 
\end{equation} 
where $ij$ are Dirac indices. The color indices of quark fields are contracted. Gauge links 
along the $n$-direction are suppressed in the above expression,  or one can eliminate them in the light-cone gauge $n\cdot G =0$.   In the above we have neglected higher-twist 
contributions and the contribution at leading twist with the structure $\gamma_5\gamma_\perp^\mu \gamma^-$. The later is called as transversity distribution in the case 
of a spin-1/2 hadron\cite{JaJi}. The distribution defined with the structure $\gamma_5\gamma_\perp^\mu \gamma^-$ will 
not contribute in our case because of its chirality.  Comparing with quark distributions 
of a spin-1/2 hadron, there is an additional quark distribution related to the tensor polarization 
of the spin-1 hadron. It is also noted that the additional distribution $\tilde q_L(x)$ in Eq.(\ref{Quark}) 
is multiplied with the same $\gamma^-$ as the unpolarized quark distribution $q(x)$. Therefore, 
the perturbative coefficient functions related to the two distributions will be the same.   
\par 
Four gluon distributions can be defined with the twist-2 gluonic operator. They are: 
\begin{eqnarray} 
 && \frac{1}{x P^+} \int\frac{d\lambda}{2\pi} e^{-ix\lambda P^+} 
\langle H \vert G^{a,+\mu}(\lambda n) G^{a,+\nu} (0) \vert H\rangle 
\nonumber\\ 
 && = - \frac{1}{2} g_\perp^{\mu\nu} \biggr ( g(x) + \tilde g_{L}(x)  \tilde S_{L} \biggr ) - \frac{1}{2} T_\perp^{\mu\nu} g_T (x) -\frac{i}{2} \epsilon_\perp^{\mu\nu} g_L(x) S_L, 
\label{Gluon} 
\end{eqnarray} 
where $G^{\mu\nu}$ is the gluon field strength tensor and the indices $\mu$ and $\nu$ are tranverse. Again, gauge links should be supplemented 
to make the definition gauge-invariant.  Comparing with gluon distributions of a spin-1/2 hadron, here we have two additional gluon distributions related to the tensor polarization. The distribution $g_T$ was first identified in \cite{JaMa,Mano} with OPE. 
\par 
The introduced parton distributions or their certain linear combinations  have the interpretation of the probabilities to find the parton with a definite polarization in the hadron. E.g., such an interpretation of $g_T$ is given in \cite{JaMa}. 
The parton distributions depend on the renormalisation scale $\mu$. The $\mu$-dependence of the most distributions is known. The $\mu$-dependence of $q(x)$ and $g(x)$, 
is governed by the standard DGLAP equation. $\tilde q_L (x)$ and $\tilde g_L (x)$ satisfy the same DGLAP equation as $q(x)$ and $g(x)$ do. 
Similarly, the $\mu$-dependence of distributions related to $S_L$ is the same as that of the correspond distributions of a spin-1/2 hadron. The $\mu$-dependence of 
$g_T(x)$ has been derived in \cite{EvoG}. We have re-derived it. The result is:
\begin{eqnarray} 
  \frac{\partial g_T (x,\mu)}{\partial \ln \mu } &=& \frac{\alpha_s}{\pi} \int_x^1 \frac{d\xi}{\xi} 
      g_T (\xi, \mu) \biggr [   \frac{ 2 N_c z}{(1-z)_+ }  + \delta (1-z) \biggr ( \frac{11}{6} N_c -\frac{1}{3} N_f \biggr ) 
        + {\mathcal O}(\alpha_s) \biggr ] , 
\end{eqnarray} 
with $z = x/\xi$.  This confirms the result in \cite{EvoG}. 

\par 
The hadronic tensor with the contributions from twist-2 operators or twist-2 distributions can be calculated in a standard way.  We will skip the detail of the calculation and only give the results here. We introduce the following quantities:
\begin{equation}
  p^\mu =(P^+,0,0,0), \ \ \ \ \hat p^\mu = p^\mu -\frac{q\cdot p}{q^2} q^\mu, \ \ \   x =\frac{Q^2}{2 p\cdot q}  = \frac{2 x_B}{ 1 +\sqrt{ 1 + 4 M^2 x_B /Q^2}}.
\end{equation} 
Here $x$ is Nachtmann variable introduced in \cite{Nacht} which takes the target mass correction from 
kinematics into account. 
For $Q^2$ large enough, one can neglect the hadron mass and has $x\approx x_B$.
Our result for the twist-2 part of the hadronic tensor takes the form:
\begin{eqnarray} 
W^{\mu\nu} \biggr\vert_{{\rm twist}-2} &=& \left (- g^{\mu\nu} + \frac{q^\mu q^\nu}{q^2} \right )  \left  ( F_1 (x_B, Q^2) + \tilde S_L \tilde F_1 (x_B,Q^2)  \right ) 
  + \frac{x}{p\cdot q} \hat p^\mu \hat p^\nu  \biggr (  F_2 (x_B,Q^2) 
\nonumber\\  
       &&                    + \tilde S_L \tilde F_2 (x_B,Q^2) \biggr  )  
  + \frac{i}{p\cdot q} S_L \epsilon^{\mu\nu\alpha\beta} p_\alpha q_\beta   \hat G_1 (x_B, Q^2) 
+ T_\perp^{\mu\nu} \hat G_T(x_B,Q^2), 
\label{WT2}        
\end{eqnarray}
and the structure functions at leading order of $\alpha_s$ are expressed of twist-2 parton distributions as: 
\begin{eqnarray} 
F_1 (x_B, Q^2) &=& \frac{1}{2} q(x), \ \ \  F_2(x_B, Q^2) = q(x), 
\ \ \ \ 
\hat G_1 (x_B, Q^2) =  q_L(x), 
\nonumber\\
\hat G_T (x_B, Q^2) &=&  \frac{\alpha_s}{4\pi} \int_x^1 d\xi \frac{x^2}{\xi^3}  g_T (\xi, \mu^2).
\label{FACT2} 
\end{eqnarray} 
In the factorization the structure functions are expressed as convolutions of perturbative coefficient functions 
with twist-2 parton distributions. For the first 5 structure functions in Eq.(\ref{WT2}) they are $\delta$-functions. 
We do not explicitly give the contributions with anti-quark distributions here and later in the case 
with twist-3 contributions. They can easily be obtained. 
The results of $\tilde F_{1,2}$  can be obtained from the results of $F_{1,2}$ by replacing $q(\xi,\mu^2)$
with $\tilde q_L(\xi,\mu^2)$, and $g(x,\mu^2)$ with $\tilde g(x,\mu^2)$ beyond the leading order.    
\par 
We have written our results in the form so that the perturbative coefficient functions 
are exactly the same appearing in the corresponding 
structure functions for DIS with a spin-1/2 hadron. Therefore, the one-loop corrections to the results 
in the first line in Eq.(\ref{FACT2}) and to $\tilde F_{1,2}$ are known.  The leading order of $\hat G_T$ is of $\alpha_s$. 
We have extracted the perturbative coefficient function of $\hat G_T$ from 
the forward scattering $\gamma^* g \to q\bar q \to \gamma^* g$. The result agrees with that in \cite{JaMa}. It should be noted that 
$\hat G_T$ only receives contributions from gluons in the target. It does not receives any 
contributions from quarks at twist-2.

\par
Now we turn to the twist-3 part of the hadronic tensor.  At twist-3 there are many different operators, 
e.g., there are twist-3 operators defined with bilinear quark fields like $\bar \psi (\lambda n) \psi (0)$  and those defined  with bilinear quark fields combined with one gluon field strength operators like
$\bar\psi(\lambda_1 n) G^{+\mu} (\lambda_2 n) \psi (0)$, or with the covariant derivative $D^\mu(\lambda_2 n) = \partial^\mu + i g_s G^\mu (\lambda_2 n)$  instead of $G^{+\mu} (\lambda_2 n)$ given in \cite{QiuSt,EFTE}. However, not all of their matrix elements are independent, as shown in \cite{TW3R,JiG2,JiG21}.  This will also be shown 
later in this work. To consistently  
factorize the relevant structure functions, one should take those operators which are independent. One should use an adequate set of indpendent operators for a given process. In our case, the twist-3 effect comes from the transverse motion of a single incoming parton 
as given in Fig.2a and from multi-parton scattering as given in Fig.2b. Therefore, it is convenient to take the operators defined with the covariant derivative, in which the part with the derivative represents the effect of the transverse motion. We define here: 
\begin{eqnarray} 
&& P^+ \int \frac{dy_1 dy_2}{(2\pi)^2}
   e^{ -iy_2 (x_2-x_1) P^+ -i y_1 x_1 P^+ }
  \langle P, \epsilon \vert
           \bar\psi (y_1n )   D^\mu (y_2n) \psi(0) \vert P,\epsilon \rangle 
\nonumber\\                      
           &&= \frac{1}{4} \biggr [ 
            \gamma^ - \biggr ( D_F (x_1,x_2) \epsilon_\perp^{\mu\nu} S_{\perp\nu} + i \tilde D_F (x_1,x_2) \tilde S^\mu_{\perp} \biggr ) 
\nonumber\\
   && \ + i \gamma_5 \gamma^- \biggr ( D_\Delta (x_1,x_2) S_\perp^\mu +  i \tilde D_\Delta (x_1,x_2) \epsilon_\perp^{\mu\nu} \tilde S_{\perp\nu} \biggr ) \biggr ]+\cdots
\label{DTW3} 
\end{eqnarray}
where $\cdots$ denote power-suppressed contributions. The definition is given in the light-cone gauge $n\cdot G=0$. 
In other gauges, gauge links need to be supplemented. 
From symmetries of QCD one can show: 
\begin{eqnarray}
&&   D_F(x_1,x_2) = -D_F (x_2,x_1), \ \ \ \ \ D_\Delta (x_1,x_2) =D_\Delta (x_2,x_1), 
\nonumber\\  
&& \tilde D_F(x_1,x_2) =  \tilde D_F (x_2,x_1), \ \ \ \ \ \tilde D_\Delta (x_1,x_2) = -\tilde D_\Delta (x_2,x_1). 
\end{eqnarray}
\par 
Similarly, there are four independent twist-3 gluon distributions defined with operators of gluon field strength tensor. Two of them 
are defined with the structure constant $f^{abc}$:    
\begin{eqnarray} 
 &&\frac{ i f^{abc} g_{\alpha\beta} }{P^+} 
      \int\frac{d y_1 d y_2}{4\pi} 
   e^{ i(-y_1 x_1  +y_2x_2)P^+}
 \langle H \vert  G^{a,+\alpha}(y_1 n) G^{b, +\mu}(0) G^{c,+\beta} (y_2 n ) \vert H  \rangle 
\nonumber\\ 
   &&  =  T_G (x_1,x_2) \epsilon_\perp^{\mu\nu} S_{\perp\nu} + i \tilde T_G (x_1,x_2) \tilde S^\mu_{\perp}.
\label{TW3G}     
\end{eqnarray}
The contributions involving these matrix elements of the twist-3 gluonic operators 
appear at higher order of $\alpha_s$. They will be not considered here. We only notice here that symmetries, like Bose symmetry of gluons,  
give constraints of the form of the two distributions. Different parameterizations  
of these twist-3 gluon distributions exist, e.g., in \cite{TG3}. 
There are another two twist-3 gluon distributions defined by replacing $if^{abc}$ in Eq.(\ref{TW3G}) with $d^{abc}$. 
These two distributions will not give contributions to DIS, because that the operator with $d^{abc}$ is odd under charge conjugation
and the product of currents in the hadronic tensor in Eq.(\ref{WT}) is $C$-even. 
The scale-dependence of twist-3 operators has been studied in \cite{ETW3-1,BMP,BB,ETW3-2}.

\begin{figure}[hbt]
\begin{center}
\includegraphics[width=9cm]{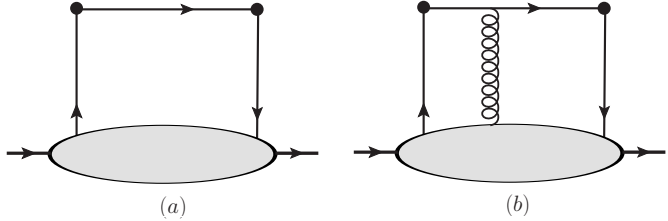}
\end{center}
\caption{The diagrams for twist-3 contributions. The black dots denote the insertion of electromagnetic current operators. A cut is implied for all diagrams.    }
\label{P1}
\end{figure}
\par
  
\par 
The contributions from these twist-3 matrix elements to the hadronic tensor at leading order of $\alpha_s$ are given by diagrams 
in Fig.2. In this letter, we work with the light-cone gauge. 
As discussed in detail in \cite{tw4}, the contributions from Fig.2a contains not only the contributions of twist-2 but also 
the contributions of twist-3 or higher twist. The twist-2 contributions from Fig.2a are those given in the first line of Eq.(\ref{FACT2}).  
The leading-power contribution of Fig.2b is of twist-3 in the gauge $n\cdot G=0$. We will take this gauge to derive our results. 
The contribution from Fig.2a reads: 
\begin{eqnarray} 
W^{\mu\nu}\biggr\vert_{2a} = \frac{1}{4\pi} \int  \frac{ d^4 k d^4 y}{(2\pi)^4} e^{-i y\cdot k} 
\langle P \vert \bar \psi (y)   H^{\mu\nu} (k,q)  \psi(0) \vert P \rangle, \ \ \ 
  H^{\mu\nu} (k,q) =  2 \pi \delta ((k+q)^2) \gamma^\mu \gamma\cdot (k+q)
    \gamma^\nu,   
\label{WL}     
\end{eqnarray}
where $k$ is the four momentum of the quark entering the scattering with the virtual photon.  
To separate the contributions of different twists, we first expand $H^{\mu\nu}(k,q)$ around the momentum $\hat k^\mu =(k^+,0,0,0)$: 
\begin{equation} 
 H(k,q) =  H(\hat k,q) + k_\perp^\rho \frac{\partial H(k,q)}{\partial k_{\perp\rho}} \biggr\vert_{k=\hat k} +\cdots. 
\label{EXPH}  
\end{equation}  
Then we decompose the quark field into the large- and small component in the high energy limit: 
\begin{equation} 
   \psi(x) = \psi_+ (x) + \psi_-(x), \ \ \ 
    \psi_+ (x) = \frac{1}{2} \gamma^- \gamma^+ \psi (x), \ \ \ \psi_-(x) =  \frac{1}{2} \gamma^+ \gamma^- \psi (x), 
\label{LSQF}     
\end{equation} 
where $\psi_+(\psi_-)$ is the large(small) component. With equation of motion one can derive: 
\begin{equation} 
2 \partial^+ \psi_- (x) = -\gamma^+\gamma_\perp \cdot D_\perp \psi_+ (x). 
\end{equation}
This indicates that the small component $\psi_-$ in comparison with the large component $\psi_+$ is power-suppressed in the matrix element 
in Eq.(\ref{WL}). One can use the expansion in Eq.(\ref{EXPH}) and the decomposition in Eq.(\ref{LSQF}) to obtain contributions 
of different twists.  
\par 
The leading twist contribution from Fig.2a is given by taking the first term in the expansion and neglecting the small 
component of quark fields in Eq.(\ref{WL}).  
The twist-3 contribution from Fig.2a has two parts. The first one is from the second term in the expansion in Eq.(\ref{EXPH}), where 
all quark fields are the large components in Eq.(\ref{WL}). This gives a twist-3 contribution involving operators 
like $\bar\psi \partial_\perp \psi$. The second part is given by taking the first term in Eq.(\ref{EXPH}), where one small component 
in Eq.(\ref{WL}) is involved. This gives another twist-3 contribution involving operators 
like $\bar\psi D_\perp \psi$.    
\par 
The twist-3 contribution from Fig.2b is obtained in a straightforward way, where one lets all partons have the momenta along the $+$-direction 
and the gluon line is for a transversely polarized gluon. One then obtains the twist-3 contribution from Fig.2b involving the operator 
$\bar\psi G_\perp \psi$. Summing this contribution with the first part of Fig.2a, we find that the sum can be written 
in a form involving the operator $\bar \psi(y_1 n) D_\perp(y_2 n) \psi (0)$.  
At the end we have the total twist-3 contribution of the hadronic tensor:  
\begin{eqnarray}
W^{\mu\nu}\biggr\vert_{{\rm twist}-3} &=&  -\frac{1}{2 p\cdot q} \biggr [  \tilde S^\mu_{\perp} \hat p^\nu 
      + \tilde S^\nu_{\perp} \hat p^\mu  \biggr ] 
  \int dy   \biggr ( 
 \tilde  D_F (x, y)  - \tilde  D_\Delta  (x, y ) \biggr )
\nonumber\\
 && \ \ -i \frac{1}{2  q ^2} \epsilon^{\mu\nu\alpha\beta} q_\alpha S_{\perp\beta}  \int  dy \biggr ( 
  D_\Delta (x, y)  +  D_F  (x, y ) \biggr ),      
\label{WT3}   
\end{eqnarray}    
with the integration range $1 > y > -1+x$.  It should be noted that the case with $y>0$ and 
that of $y<0$  represent different partonic processes. For $y<0$, the process is the forward scattering of $\gamma^* q\bar q \to \gamma^* g$, while for $y>0$ the partonic process is $\gamma^* g q \to \gamma^* q$. This can be seen clearly by using multi-parton states as a target 
to calculate the hadronic tensor and twist-3 matrix elements, similarly to the study of single spin asymmetries in \cite{MPS1}, where  the factorization involves  twist-3 operators.   
\par 
Comparing DIS with a spin-1/2 hadron, the last term in Eq.(\ref{WT3})  corresponds to the structure function related to the transverse polarization of the spin-1/2 hadron. This structure function 
has been studied in detail in \cite{JiG2,JiG21}, where one-loop correction has been obtained. Therefore, the one-loop correction to the last term  in Eq.(\ref{WT3}) is also known.  It may be possible to extract one-loop correction to the first term 
in Eq.(\ref{WT3}) from the results of \cite{JiG2,JiG21}. 
\par 
It is interesting to realize that the special form of the tree-level result  in Eq.(\ref{WT3}) can be written in a more simple form by using equation of motion of QCD, as discussed in the case of a spin-1/2 hadron in \cite{JiG2}.  For this we define the following two twist-3 functions 
with the mentioned bilinear quark operator: 
% redefind qt! 
\begin{eqnarray} 
   x P^+ \int\frac{d\lambda}{2\pi} e^{-ix\lambda P^+} 
  \langle H \vert \bar \psi (\lambda n)  \gamma_5 \gamma_\perp^\mu  \psi  (0) \vert H \rangle 
  & =& q_T (x) S_\perp^\mu, 
\nonumber\\
x P^+\int\frac{d\lambda}{2\pi} e^{-ix\lambda P^+} 
  \langle H \vert \bar \psi (\lambda n) \gamma^\mu_\perp   \psi  (0) \vert H \rangle & =& 
    \tilde q_T (x) \tilde S_{\perp}^\mu. 
\end{eqnarray}
One can find the following operator-identities and the hadronic tensor: 
\begin{eqnarray}
q_T (x) &=&  \int d y \biggr ( D_F (x,y) + D_\Delta (x,y) \biggr ) , \quad\quad  
 \tilde q_T (x) =- \int  dy  \biggr ( \tilde D_F (x,y) - \tilde  D_\Delta (x,y) \biggr ) , 
\nonumber\\
W^{\mu\nu}\biggr\vert_{{\rm twist} -3}  &=&  \frac{1 }{2 p\cdot q} \left ( \tilde S^\mu_{\perp} \hat p^\nu +\tilde S^\nu_{\perp} \hat p^\mu \right ) \tilde q_T (x)  -i \frac{1 }{2  q ^2}   q_T (x) \epsilon^{\mu\nu\alpha\beta} q_\alpha S_{\perp\beta}. 
\label{TW3TR}    
\end{eqnarray}  
However, for factorization one should use the result in Eq.(\ref{WT3}) for higher-order corrections and study of $Q^2$-dependence
as discussed in \cite{JiG2}. In this work we do not introduce new notations of structure functions for the twist-3 part. In the below, we will 
use the notation of our  tree-level results in Eq.(\ref{TW3TR}) for differential cross-sections. It should be kept in mind 
that beyond tree-level the structure functions of the twist-3 part should be written as convolutions with those twist-3 matrix elements 
in Eq.(\ref{DTW3}). Our results in Eq.(\ref{WT2}) and Eq.(\ref{TW3TR}) are covariant as discussed before about the light-cone frame,  
and are $U_{em}(1)$-gauge invariant.       
\par 
With our results of structure functions we can obtain the differential cross-section for arbitrary $\vec m$ 
with $\lambda_m =0,\pm 1$: 
\begin{eqnarray}    
\frac{ d \sigma (\lambda_m) }{dx_B dy d\phi   }  &=&  \frac{2 y \alpha^2}{Q^2} 
\biggr [ F_1 (x_B,Q^2) + a_m \tilde F_1 (x_B,Q^2)   +\frac{1-y}{y^2}  \left ( F_2 (x_B,Q^2) 
  +  a_m  \tilde F_2 (x_B,Q^2) \right )
\nonumber\\
   && - \frac{1-y}{2y^2}b_m \sin^2 \theta_m \hat G_T (x_B,Q^2) \cos 2\phi - \lambda \lambda_m \frac{2-y}{y}  \cos\theta_m 
   \hat G_1 (x_B,Q^2)        
\nonumber\\
   && - \frac{\sqrt{1-y}}{2 y^2 Q} \cos\phi \biggr  (  - (2-y)  
    b_m\sin 2\theta_m    \tilde q_T (x)    
  - 2\lambda \lambda_m  y  \sin\theta_m    q_T (x)  \biggr )  \biggr ],  
\label{dsigma}   
\end{eqnarray} 
where $a_m$ and $b_m$ are parameters of $\lambda_m$. They are:
\begin{equation} 
  b_m = 3 \vert \lambda_m \vert -2,\quad\quad a_m =\frac{1}{4} b_m (1- 3\cos^2 \theta_m). 
\end{equation}   
Eq.(\ref{dsigma}) is our main result. The differential cross-section is given in the laboratory frame, the azimuthal angle $\phi$ is defined in Fig.1. 
The terms in the first two lines are of leading-twist part of $W^{\mu\nu}$. They are also 
at the leading power of $1/Q$ in the differential cross-section. The terms in the last line are from the twist-3 part of $W^{\mu\nu}$, 
they are suppressed by $1/Q$. 
The result is at the accuracy of the next-to-leading power of $1/Q$. In Eq.(\ref{dsigma}) we have neglected the effect of finite target mass 
in kinematics except the effect in the relation between $x$ and $x_B$.   
From the result one can see that the tensor-polarized gluon distribution $g_T$ in Eq.(\ref{Gluon}) characterizes a $\cos 2\phi$-dependence 
at leading power of $1/Q$. The twist-3 matrix element $\tilde q_T$ involving tensor polarization of $H$ leads to a $\cos\phi$-dependence. 
\par 
Some terms vanish if one takes $\theta_m =90^\circ$ in Eq.(\ref{dsigma}), e.g., the contributions of $\hat G_1$ and $\tilde q_T$. 
In fact these contributions do not vanish, they are kinematically suppressed by an overall factor $M/Q$. To see this we give 
the differential cross-section in the case of $\theta_m =90^\circ$:
\begin{eqnarray}    
\frac{ d \sigma (\lambda_m) }{dx_B dy d\phi   }  &=&  \frac{2 y \alpha^2}{Q^2} 
\biggr [ F_1 (x_B,Q^2) + \frac{b_m}{4}  \tilde F_1 (x_B,Q^2)   +\frac{1-y}{y^2}  \left ( F_2 (x_B,Q^2) 
  +  \frac{b_m}{4}   \tilde F_2 (x_B,Q^2) \right )
\nonumber\\
   && - \frac{1}{2 y^2 }b_m (1-y)  \hat G_T (x_B,Q^2) \cos 2\phi  + \lambda \lambda_m  \frac{\sqrt{1-y}}{ y Q}     q_T (x)   \cos\phi 
\nonumber\\      
  && - (2-y) \frac{2 M x_B\sqrt{1-y}}{ yQ} \biggr (   
  \lambda \lambda_m  \hat G_1 (x_B,Q^2)         
   - b_m \frac{\sqrt{1-y}}{ y Q}     \tilde q_T (x)    
     \biggr )  \cos\phi \biggr ]. 
\label{dsigma1}   
\end{eqnarray}    
From the above equation one can see that in the case of $\theta_m =90^\circ$, the contribution from $\hat G_1$ becomes 
power-suppressed and is $\cos\phi$-dependent. The contribution from $\tilde q_T$ is now at order of $1/Q^4$. In comparison 
with Eq.(\ref{dsigma}), the contribution from $\hat G_1$ and $\tilde q_T$ are suppressed by the overall 
factor $M/Q$ in the case $\theta_m=90^\circ$. 
This factor is purely from kinematics. The effect from higher twist in $\hat G_1$ or higher-twist correction to contribution of $\tilde q_T$ 
will also be suppressed by the same kinematic factor and are not at leading power of $Q$. From Eq.(\ref{dsigma1})  the contributions from $\hat G_1$ and $q_T$ give 
a $\cos\phi$-dependence at the same order of $1/Q$, although they are of different twist.     
\par 
Another special case is with $\theta_m=0^\circ$. In this case we have:
\begin{eqnarray}    
\frac{ d \sigma (\lambda_m) }{dx_B dy d\phi   }  &=&  \frac{2 y \alpha^2}{Q^2} 
\biggr [ F_1 (x_B,Q^2)- \frac{b_m}{2}  \tilde F_1 (x_B,Q^2)   +\frac{1-y}{y^2}  \left ( F_2 (x_B,Q^2) 
  -  \frac{b_m}{2}   \tilde F_2 (x_B,Q^2) \right )
\nonumber\\
   && -b_m \frac{ 2 M^2 x_B^2 (1-y)^2}{ y^2 Q^2}  \hat G_T (x_B,Q^2)   - \lambda \lambda_m \frac{2-y}{y}    \hat G_1 (x_B,Q^2)         
\nonumber\\
   && + \frac{ 2M x_B (1-y)}{ y^2 Q^2}  \biggr  (  - (2-y)  
    b_m    \tilde q_T (x)    
  - \lambda \lambda_m  y     q_T (x)  \biggr )  \biggr ]. 
\label{dsigma2}   
\end{eqnarray}
There is no azimuthal dependence as expected. Here, the contributions from $\hat G_T$ and the twist-3 part of $W^{\mu\nu}$ 
are suppressed by the factor $M/Q$ from kinematics, as in the case discussed before. 

\par 
Experimentally, the azimuthal dependence in DIS with a spin-1 nucleus can be studied with fixed target experiment of J-Lab and at the proposed EIC\cite{EIC} . It will be interesting to see evidences of the existence of those structure functions related 
to the tensor polarization. They are $\tilde F_{1,2}$, $\hat G_T$ and $\tilde q_T$. If there is no interaction between nucleons in a nucleus, 
they are zero. Hence,  experimental study of the azimuthal dependence in DIS with a spin-1 polarized nucleus 
will provide more information about the inner structure of the nucleus. The experimental results in \cite{b1hermes} indicates that the structure function called $b_1$ according to \cite{HJM} 
can be nonzero. $b_1$ is related to the tensor polarization $\tilde S_L$ and is equivalent 
to the defined structure function $\tilde F_1$ or $\tilde F_2$ here. From our result, the structure 
function $\tilde F_{1,2}$ gives a constant contribution 
in the azimuthal distribution.

\par 
Among the structure functions related to tensor polarization, $\hat G_T$ is of the most interesting, as discussed in \cite{JaMa}. 
The existence of this structure function will clearly indicate that a nucleus has a nonzero gluon content which can not be interpreted 
with the gluon content of each individual nucleon in the nucleus. From Eq.(\ref{FACT2}), the contribution of $g_T$ to the differential cross 
section is suppressed by $\alpha_s$ relatively to other structure functions. It may be easier to extract $g_T$ from DIS with the restriction 
that the final state consists a heavy quark $Q\bar Q$-pair with other possible hadrons. Then, all parton distributions contribute at the same order 
of $\alpha_s$. The results for the first three structure functions in Eq.(\ref{FACT2}), hence also for $\tilde F_{1,2}$, can be found 
in literature about DIS with a spin-1/2 target. We denote $\hat G_T$ in this case as $\hat G_{T,Q\bar Q}$ and derive here the result 
for $\hat G_{T,Q\bar Q}$ by considering 
the forward scattering of $\gamma^* g \to Q\bar Q \to \gamma^* g$:    
\begin{eqnarray} 
\hat G_{T,Q\bar Q} (x_B, Q^2) &=& \frac{\alpha_s}{4\pi}\int^1_{x/(1-4 x_m)} \frac{d\xi}{\xi} g_T (\xi) \left [ \beta (z^2+ 2 x_m (1-z) ) +4 x_m (z+x_m) \ln\frac{1+\beta}{1-\beta} \right ] + {\mathcal O} (\alpha_s^2), 
\nonumber\\
    && z = \frac{x}{\xi}, \ \ \ x_m = \frac{m_Q^2}{2 p\cdot q}, \quad\quad \beta^2= 1 -\frac{4 x_m}{1-z}. 
\end{eqnarray}
In the above $m_Q$ is the mass of the heavy quark $Q$.  In the limit of $m_Q \to 0$, we find from $\hat G_{T,Q\bar Q}$ the result of $\hat G_T$ in Eq.(\ref{FACT2}).   
\par

\par 
To summarize: We have studied the azimuthal dependence of DIS 
with a spin-1 hadron or nucleus, where the initial hadron is polarized in an arbitrary direction. 
In the first step we have re-derived the twist-2 contributions and the evolution of the tensor-polarized gluon distribution. 
The results agree with existing one.  
We then have derived the twist-3 contributions at tree-level. There are two structure functions 
from twist-3 operators. One is related to the tensor polarization, another is related to the vector polarization. 
The azimuthal distribution is given for the case that the initial hadron is polarized in an arbitrary direction. 
In general structure functions related to the tensor polarization of a spin-1 nucleus will be zero, if nucleons
do not interact or  are not correlated inside the nucleus. Therefore, experimental study of DIS with a spin-1 nucleus will provide information 
about the interaction or the correlation. Our result shows that these structure functions can be extracted from the studied azimuthal 
distribution.

\par\vskip20pt
\noindent
{\bf Acknowledgments}
\par
The work of J.P. Ma is supported by National Nature
Science Foundation of P.R. China(No.11021092, 11275244).  

\par

\end{document}